\documentclass[aps,showpacs,prd,twocolumn]{revtex4}%
\usepackage{amssymb}
\usepackage{amsfonts}
\usepackage{amsmath}
\usepackage{graphicx}
\usepackage{color}
\usepackage[dvips]{epsfig}
\usepackage[dvips]{graphicx}
\usepackage{float} 
\begin{document}
\title{Uncharged compactlike and fractional Lorentz-violating BPS vortices in the
CPT-even sector of the standard model extension}
\author{C. Miller,$^{1}$ R. Casana,$^{1}$ M. M. Ferreira Jr.$^{1}$}
\author{E. da Hora$^{2}$}
\affiliation{$^{1}$Departamento de F\'{\i}sica, Universidade Federal do Maranh\~{a}o,
65085-580, S\~{a}o Lu\'{\i}s, Maranh\~{a}o, Brazil}
\affiliation{$^{2}$Departamento de F\'{\i}sica, Universidade Federal da Para\'{\i}ba,
58051-900, Jo\~{a}o Pessoa, Para\'{\i}ba, Brazil}

\begin{abstract}
We have investigated and verified the existence of stable uncharged Bogomol'nyi-Prasad-Sommerfeld (BPS)
vortices in the framework of an Abelian Maxwell-Higgs model supplemented
with CPT-even and Lorentz-violating (LV) terms belonging to the gauge and
Higgs sectors of the standard model extension. The analysis is performed in
two situations: first, one by considering the Lorentz violation only in the
gauge sector and then in both gauge and Higgs sectors. In the first case, it
is observed that the model supports vortices somehow equivalent to the ones
appearing in a dielectric medium. The Lorentz violation controls the radial
extension (core of the solution) and the magnetic field amplitude of the
Abrikosov-Nielsen-Olesen vortices, yielding compactlike defects in an
alternative and simpler way than that of $k-$field models. At the end, we consider the
Lorentz-violating terms in the gauge and Higgs sectors. It is shown that the
full model also supports compactlike uncharged BPS vortices in a modified
vacuum, but this time there are two LV parameters controlling the defect
structure. Moreover, an interesting novelty is introduced by the
LV-Higgs sector: fractional vortex solutions.

\end{abstract}

\maketitle

\section{\bigskip Introduction}

The investigation of stable vortex configurations has been an issue of
permanent interest since the pioneering proposal of Abrikosov-Nielsen-Olesen
(ANO) \cite{ANO}. In the early 1990s, vortex configurations were analyzed in
the context of planar theories including the Chern-Simons term \cite{CS},
which provided the possibility of having charged vortices \cite{CSV}. The
Chern-Simons vortex configurations support Bogomol'nyi-Prasad-Sommerfeld (BPS) solutions and present important connections with the physics of anyons and
the fractional quantum Hall effect \cite{Ezawa}. The Chern-Simons vortices
were studied with nonminimal coupling \cite{CSV1}, and they remain a topic of
intensive investigation with recent developments \cite{CSV2,Bolog}.
Generalized Chern-Simons vortex solutions were recently examined in the
presence of a noncanonical kinetic term \cite{Hora1} and \emph{k}-field terms (high-order derivative terms) \cite{Compact1}. Generalized vortex solutions were
also attained in the contexts of the Abelian Maxwell-Higgs (AMH) model \cite{Hora1b}
and twinlike models \cite{HoraTwin}. The \emph{k}-field theories work with nonlinear
functions of the usual kinetic term to obtain new solutions for nonlinear
systems, with interesting applications in cosmology and inflation
\cite{Cosmology}, dark matter \cite{DM}, tachyon matter \cite{Tachyon}, ghost
condensates \cite{GC} and topological defects \cite{kfield}. Concerning
topological defects, the higher-order kinetic terms engender the formation of
\emph{k}-defects (compactlike solutions), structures whose core can be much smaller
than the one of the usual solutions \cite{kfield,Hora2,Hora3}. When a \emph{k}-defect
presents a compact support, it is dubbed a compacton \cite{Compact2,Compact3}.
In the present work, we show that the inclusion into the AMH model of Lorentz-violating (LV) terms-belonging to the theoretical framework of the
standard model extension (SME)- can yield compactlike vortex solutions and
also fractional quantization of the magnetic flux.

Lorentz symmetry violation has been much investigated in the past few years,
having as a theoretical framework the standard model extension \cite{Colladay}%
, based on the idea of a spontaneous Lorentz-symmetry breaking in a theory
defined at the Planck scale \cite{Samuel}\textbf{. } The SME incorporates
LV terms, generated as non-null vacuum expectation values
of Lorentz tensors, in all sectors of the standard model. The investigations
in the context of the SME concern mainly the fermion sector
\cite{Fermion,Fermion2,Fermion3}, extensions involving gravity \cite{Gravity}%
,\cite{Gravity2}, and the gauge sector \cite{Jackiw}-\cite{Klink3}. The gauge
sector of the SME is composed of a CPT-odd part (the Carroll-Field-Jackiw term
\cite{Jackiw}) and a CPT-even part, which is represented by the tensor
$\left(  k_{F}\right)  _{\mu\nu\alpha\beta}.$ The electrodynamics modified by
this term has been investigated since 2002, with a twofold purpose: to
scrutinize the new physical properties induced by its 19 Lorentz-violating
coefficients and to impose tight upper bounds on the magnitude of these
coefficients. \ The CPT-even tensor $\left(  k_{F}\right)  _{\mu\nu\alpha
\beta}$ has the same symmetries as the Riemann tensor, and a double null
trace, $\left(  k_{F}\right)  ^{\mu\nu}{}_{\mu\nu}=0$. References. \cite{KM1,KM2}%
, stipulated the existence of 10 components sensitive to birefringence
and 9 that are called nonbirefringent. The 10 birefringent components are
severely constrained to the level of $1$ part in $10^{32}$ by
spectropolarimetry data of cosmological sources \cite{KM1,KM2}. The
nonbirefringent coefficients are constrained by other tests, involving the
study of Cherenkov radiation \cite{Cherenkov2}, the absence of emission of
Cherenkov radiation by ultrahigh energy cosmic rays \cite{Klink2,Klink3}, and
the subleading birefringent behavior of the nonbirefringent parameters
\cite{Qasem}, able to yield upper bounds of up to $1$ part in $10^{37}$ on
these coefficients. The gauge sector of the SME has also been investigated in
the context of arbitrary dimensional operators \cite{Kostelec}\textbf{.}

Effects of Lorentz violation on topological defects have been investigated in
distinct scenarios. Some works have examined the role played by
Lorentz-violating terms on defects defined in the framework of scalar systems
\cite{Defects}, revealing the associated properties and preservation of the
linear stability. In another line, the existence of monopole solutions in the
presence of the Lorentz-violating Carroll-Field-Jackiw term was first studied
in Ref. \cite{Monopole1}. Recently, the existence of monopoles in the
framework of a rank-2 antisymmetric Kalb-Ramond tensor field, generated in a
spontaneous symmetry breaking, was analyzed in Ref. \cite{Monopole2},
unveiling some similarities between the profiles of the antisymmetric
Kalb-Ramond monopole and the usual O(3) one. A more complete study of
topological defects in the context of field theories endowed with a tensor
which spontaneously breaks Lorentz symmetry was accomplished in Ref.
\cite{Seifert}. Up the moment, there is no report of works
investigating\ vortex solutions in the presence of Lorentz-violating terms,
except for a preliminary contribution \cite{Baeta2}. There are some
addressing vortex configurations in noncommutative scenarios which yield
Lorentz symmetry only as a by-product \cite{Gamboa}.

In this work, we
investigate for the first time the formation of stable uncharged vortex configurations in the context of a Lorentz-violating and CPT-even AMH
electrodynamics in two situations: (i) with a Lorentz-violating term only in
the gauge sector, and (ii) introducing Lorentz-violating terms simultaneously
in the gauge and Higgs sectors of the model. In both cases the Higgs sector is
endowed with a particular and appropriated fourth-order self-interacting
potential. In the first case, one verifies the existence of BPS solutions,
governed by analogue equations to the AMH model. The Lorentz-violating
parameter appears as a key element for defining an effective electrical
coupling constant and modifying the mass of the boson fields. The vortex
profiles, generated by numerical methods, reveal that the Lorentz- violating
parameter acts as an element able to control the radial extension of the
defect (vortex core), in a similar way as observed in \emph{k}-field theories which
engender compactlike structures. Finally, in order to address a more complete
Lorentz-violating framework supporting vortex solutions, we consider the AMH
model with two CPT-even Lorentz-violating terms in the Higgs sector as well.
We achieve the equations of motion and evaluate the fourth-order potential
(compatible with the BPS solutions), which entails two vacua. After showing that the Higgs LV
parameter induces energy instability, we find the
self-interacting potential (endowed with only one vacuum) and BPS equations
for the stable uncharged vortices. We finally demonstrate that the asymptotic
solutions are only compatible with a modified vortex ansatz that yields
fractional magnetic flux quantization.

\section{The theoretical framework}

The basic framework of our investigation is a CPT-even and Lorentz-violating
AMH model. The proposal consists in supplementing the usual Maxwell-Higgs
Lagrangian (that provides the ANO solutions) with the CPT-even terms belonging
to the structure of the\ standard model extension, that is,
\begin{align}
\mathcal{L}_{1+3}  &  =\mathcal{-}\frac{1}{4}F_{\mu\nu}F^{\mu\nu}-\frac{1}%
{2}\kappa^{\rho\alpha}F_{\rho\sigma}F_{\alpha}^{\text{ \ }\sigma
}\nonumber\\[0.2cm]
&  +\left\vert \mathcal{D}_{\mu}\phi\right\vert ^{2}+\left(  k_{\phi\phi
}\right)  ^{\mu\nu}\left(  \mathcal{D}_{\mu}\phi\right) ^{\ast}\left(
\mathcal{D}_{\nu}\phi\right) \label{L2}\\[0.2cm]
&  -\frac{1}{2}\left(  k_{\phi F}\right)  ^{\mu\nu}F_{\mu\nu}\left\vert
\phi\right\vert ^{2}-U\left(  \left\vert \phi\right\vert \right)  .\nonumber
\end{align}
Here, $\kappa^{\mu\nu}=\left(  k_{F}\right)  ^{\mu\alpha\nu}{}_{\alpha}$ is a
traceless tensor containing the nine nonbirefringent components of the
CPT-even gauge sector \cite{Altschul}, $\left(  k_{\phi\phi}\right)  ^{\mu\nu};
$ and $\left(  k_{\phi F}\right)  ^{\mu\nu}$ are dimensionless real symmetric
and antisymmetric tensors, respectively, representing the complete Abelian
Lorentz-violating and CPT-even Higgs sector of the SME \cite{Colladay}. Note
that $\left(  K_{F}\right)  ^{\mu\alpha\nu\beta}{}$ is the CPT-even tensor
that encloses 19 components, 10 birefringent and 9 nonbirefringent
\cite{KM1,KM2}. The term $D_{\mu}\phi=\partial_{\mu}%
\phi-ieA_{\mu}\phi$ is the usual covariant derivative, $e$ is the
electromagnetic coupling constant, and $U\left(  \left\vert \phi\right\vert
\right)  $ is a fourth-order self-interaction scalar potential suitable for yielding BPS equations.

The equations of motion for the full system are
\begin{align}
&  \left.  \partial_{\nu}F^{\nu\mu}+\kappa^{\mu\alpha}\partial_{\nu}F_{\text{
}\alpha}^{\nu}-\kappa^{\nu\alpha}\partial_{\nu}F_{\text{ \ }\alpha}^{\mu
}\right. \nonumber\\[0.2cm]
&  \left.  \hspace{1cm}+\left(  k_{\phi F}\right)  ^{\nu\mu}\partial_{\nu
}(\phi^{\ast}\phi)=eJ^{\mu}+e\left(  k_{\phi\phi}\right)  ^{\mu\alpha
}J_{\alpha},\right.  \label{Ngauge1}%
\end{align}%
\begin{equation}
\mathcal{D}_{\mu}\mathcal{D}^{\mu}\phi+\left(  k_{\phi\phi}\right)
^{\mu\alpha}\mathcal{D}_{\mu}\mathcal{D}_{\alpha}\phi+\frac{1}{2}\left(
k_{\phi F}\right)  ^{\mu\nu}F_{\mu\nu}\phi+\frac{\partial U}{\partial
\phi^{\ast}}=0, \label{NHiggs1}%
\end{equation}
where the current $J_{\alpha}$ is given by
\begin{equation}
J_{\mu}=i[\phi\left(  \mathcal{D}_{\mu}\phi\right)  ^{\ast}-\phi^{\ast}\mathcal{D}_{\mu}\phi].
\label{c1a}%
\end{equation}

In the stationary regime, the Gauss's law is given by
\begin{align}
&  \left.  \left[  \left(  1+\kappa_{00}\right)  \delta_{ij}-\kappa
_{ij}\right]  \partial_{i}\partial_{j}A_{0}\right. \nonumber\\[0.12in]
&  \left.  \hspace{1cm}+\epsilon_{ija}\kappa_{0i}\partial_{j}B_{a}-\left(
k_{\phi F}\right)  _{0j}\partial_{j}(\phi^{\ast}\phi)=e\mathcal{J}%
_{0},\right.  \label{NGauss1}%
\end{align}
where
\begin{equation}
\mathcal{J}_{0}=2e\left[  1+\left(  k_{\phi\phi}\right)  _{00}\right]
A_{0}\left\vert \phi\right\vert ^{2}+\left(  k_{\phi\phi}\right)  _{0i}J_{i}.
\end{equation}
It reveals that uncharged solutions now require $\kappa_{0i}=\left(  k_{\phi
F}\right)  _{0j}=0,$ and $\left(  k_{\phi\phi}\right)  _{0i}=0$, conditions
that decouple the electric and magnetic sectors. The Amp\`{e}re's law is%
\begin{align}
&  \left.  \left(  \epsilon_{jbc}-\epsilon_{jac\ }\kappa_{ab}-\kappa
_{ja}\epsilon_{abc}\right)  \!\partial_{b}B_{c}-\kappa_{0i}\partial
_{j}\partial_{j}A_{0}\right. \nonumber\\[0.12in]
&  \left.  \hspace{1cm}+\kappa_{0j}\partial_{j}\partial_{i}A_{0}+\left(
k_{\phi F}\right)  _{12}\epsilon_{ij}\partial_{j}\left[  \phi^{\ast}%
\phi\right]  =e\mathcal{J}_{i}.\right.  \label{NAmpere1}%
\end{align}
with%
\begin{equation}
\mathcal{J}_{i}=J_{i}-\left(  k_{\phi\phi}\right)  _{ij}J_{j}+\left(
k_{\phi\phi}\right)  _{0i}J_{0}\mathbf{.\ }%
\end{equation}
On the other hand, the Higgs equation is
\begin{align}
&  \left.  \left[  \delta_{ij}-\left(  k_{\phi\phi}\right)  _{ij}\right]
\mathcal{D}_{i}\mathcal{D}_{j}\phi-ie\left(  k_{\phi\phi}\right)  _{0j}%
A_{0}\mathcal{D}_{j}\phi\right. \nonumber\\[0.12in]
&  \left.  -ie\left(  k_{\phi\phi}\right)  _{0j}\mathcal{D}_{j}\left(
A_{0}\phi\right)  +e^{2}\left[  1+\left(  k_{\phi\phi}\right)  _{00}\right]
A_{0}^{2}\phi\right. \label{NHiggs2}\\[0.12in]
&  \left.  -\left(  k_{\phi F}\right)  _{0i}\phi\partial_{i}A_{0}-\left(
k_{\phi F}\right)  _{12}B\phi-\frac{\partial U}{\partial\phi^{\ast}}=0.\right.
\nonumber
\end{align}
In the sequel we will particularize this theoretical model in two situations
of interest for studying vortex solutions.

\section{Compactlike uncharged BPS vortices in a simpler Lorentz-violating AMH
model}

We first consider an AMH model in which the Lorentz violation is only
represented by the nonbirefringent CPT-even gauge term of the SME while the
Higgs sector is supposed unaffected by Lorentz-violating terms, $\left(
k_{\phi\phi}\right)  _{\mu\nu}=0$, $\left(  k_{\phi F}\right)  ^{\mu\nu}=0.$
Hence, the model (\ref{L2}) is reduced to the form,
\begin{equation}
\mathcal{L}=-\frac{1}{4}F_{\alpha\beta}F^{\alpha\beta}-\frac{1}{2}\kappa
^{\rho\alpha}F_{\rho\sigma}F_{\alpha}{}^{\sigma}+\left\vert \mathcal{D}_{\mu}%
\phi\right\vert ^{2}-U\left(  \left\vert \phi\right\vert \right)  . \label{L1}%
\end{equation}
For this case, the fourth-order self-interaction scalar potential, $U\left(
\left\vert \phi\right\vert \right)  $, compatible with BPS solutions is
\begin{equation}
U\left(  \left\vert \phi\right\vert \right)  =\frac{e^{2}}{2\left(
1-s\right)  }\left(  v^{2}-\left\vert \phi\right\vert ^{2}\right)  ^{2},
\label{Pot}%
\end{equation}
where $s=\mbox{tr}\left(  \kappa_{ij}\right)  $ and $v$ plays the role of the
vacuum expectation value of the scalar field.

Considering the corresponding stationary Gauss's law, the condition
$\kappa_{0i}=0$ is the one that decouples the electric and magnetic sectors
(appropriate to achieving uncharged vortex solutions). With it, Eq.
(\ref{NGauss1}) is reduced to
\begin{equation}
\left[  \left(  1+\kappa_{00}\right)  \delta_{ab}-\kappa_{ab}\right]
\partial_{a}\partial_{b}A_{0}{}=2e^{2}A_{0}\left\vert \phi\right\vert ^{2}.
\label{A01}%
\end{equation}
An uncharged vortex has null electric field, being compatible with the
temporal gauge, $A_{0}=0$, for which\ the Gauss's law is trivially fulfilled.
Further, in such a gauge the modified stationary Ampere's law becomes%
\begin{equation}
\hspace{-0.2cm}\left(  \epsilon_{jbc}-\epsilon_{jac\ }\kappa_{ab}-\kappa
_{ja}\epsilon_{abc}\right)  \!\partial_{b}B_{c}=eJ_{i}, \label{B2}%
\end{equation}
where we have used $F_{ij}=\epsilon_{ijk}B_{k},$ $F_{0i}=E^{i}$.

On the other hand, the stationary equation for the complex scalar field is
\begin{equation}
\nabla^{2}\phi-i2eA_{j}\partial_{j}\phi-e^{2}A_{j}^{2}\phi+\frac{e^{2}}%
{1-s}\phi\left(  v^{2}-\left\vert \phi\right\vert ^{2}\right)  =0.
\label{Phi2}%
\end{equation}

The stationary canonical energy density in temporal gauge $\left(
A_{0}=0\right)  $ takes the form
\begin{align}
\mathcal{E}  &  =\frac{1}{2}\left[  \left(  1-s\right)  \delta_{ab}%
+\kappa_{ab}\right]  B_{a}B_{b}\nonumber\\[-0.3cm]
& \label{H}\\
&  +\left\vert \mathcal{D}_{k}\phi\right\vert ^{2}+\frac{e^{2}}{2\left(  1-s\right)
}\left(  v^{2}-\left\vert \phi\right\vert ^{2}\right)  ^{2}.\nonumber
\end{align}
As it will be clear in the next section, this situation is compatible with the
existence of ANO-like vortices in the framework of Lorentz-violating field theories.

\subsection{Uncharged vortex configurations}

In order to search for stable vortex configurations, we work in cylindrical
coordinates $\left(  r,\theta,z\right)  $, and state the usual ansatz for
static rotationally symmetric vortex solutions, with the fields parametrized
as
\begin{equation}
A_{\theta}=-\frac{a\left(  r\right)  -n}{er},~\ \ \phi=vg\left(  r\right)
e^{in\theta}, \label{Ans2}%
\end{equation}
where $a\left(  r\right)  $, $g\left(  r\right)  $ are regular scalar
functions at $r=0$ (such that the fields $A_{\theta}$ and $\phi$ are finite)
satisfying the following boundary conditions:
\begin{equation}
g\left(  r\rightarrow0\right)  \rightarrow0\;,\;\;a\left(  r\rightarrow
0\right)  \rightarrow n, \label{bc1}%
\end{equation}
and $n$ is the winding number of the topological solution. In this ansatz the
magnetic field is aligned with the $z$ axis, $B=\left(  0,0,B\left(  r\right)
\right)  $, and it holds
\begin{equation}
B\left(  r\right)  =-\frac{a^{\prime}}{er}. \label{Brot}%
\end{equation}

By considering the ansatz (\ref{Ans2}), we then rewrite Eqs. (\ref{B2}) and%
(\ref{Phi2}), attaining the following system of differential equations:
\begin{equation}
\left(  1-s\right)  B^{\prime}+2ev^{2}\frac{ag^{2}}{r} =0, \label{DE1}%
\end{equation}
\vspace{-0.5cm}
\begin{equation}
g^{\prime\prime}+\frac{g^{\prime}}{r}-\frac{1}{r^{2}}a^{2}g+\frac{e^{2}v^{2}%
}{1-s}g\left(  1-g^{2}\right)  =0, \label{DE2}%
\end{equation}
where
\begin{equation}
s=\kappa_{rr}+\kappa_{\theta\theta},
\end{equation}
is the parity-even parameter controlling the Lorentz-violating effects. It is
important to note that the introduction of the ansatz (\ref{Ans2}) produces
equations dependent only on $r$, with no reference to the \emph{z} dimension anymore.
In this sense, Eqs. (\ref{DE1}) and (\ref{DE2}) effectively describe the
physics of a planar system.

In order to obtain BPS solutions, we should
search for a set of first-order differential equations that describe the
dynamics of the system. These first-order equations are found by writing the
energy of the system as a sum of squared terms and requiring its
minimization. By using the ansatz (\ref{Ans2}) and Eq. (\ref{Brot}) in Eq.
(\ref{H}), the resulting energy density for the uncharged vortex is
\begin{align}
\mathcal{E}  &  =\frac{1}{2}\left(  1-s\right)  \left[  \frac{a^{\prime}}%
{er}\pm\frac{ev^{2}}{1-s}\left(  1-g^{2}\right)  \right]  ^{2}\nonumber\\
& \label{H3}\\[-0.2cm]
&  +v^{2}\left[  g^{\prime}\mp\frac{ag}{r}\right]  ^{2}\mp v^{2}%
\frac{a^{\prime}}{r}\pm v^{2}\frac{\left(  ag^{2}\right)  ^{\prime}}%
{r}.\nonumber
\end{align}
The squared brackets in (\ref{H3}) yield the wanted BPS equations
\begin{equation}
g^{\prime} =\pm\frac{ag}{r}, \label{BPS2}%
\end{equation}
\vspace{-0.25cm}
\begin{equation}
-\frac{a^{\prime}}{er} =B=\pm\frac{ev^{2}}{1-s}\left(  1-g^{2}\right)  .
\label{BPS1}%
\end{equation}
Under BPS equations the energy density (\ref{H3})\ reads%
\begin{equation}
\mathcal{E}_{BPS}=\mp v^{2}\frac{a^{\prime}}{r}\pm v^{2}\frac{\left(
ag^{2}\right)  ^{\prime}}{r}, \label{ED_BPS1}%
\end{equation}
whose integration under the boundary conditions,
\begin{equation}
g\left(  r\rightarrow\infty\right)  \rightarrow1,\text{ \ \ }a\left(
r\rightarrow\infty\right)  \rightarrow0, \label{bc2}%
\end{equation}
leads to topological vortex solutions possessing a finite total BPS energy,
\begin{equation}
E_{BPS}=\pm\left(  2\pi v^{2}\right)  n=ev^{2}\left\vert \Phi_{B}\right\vert .
\label{Hmin}%
\end{equation}
Here, $\Phi_{B}$\ is the magnetic flux associated with the vortex
\begin{equation}
\Phi_{B}=\int B\left(  r\right)  d^{2}\mathbf{r}=\frac{2\pi}{e}n. \label{ed}%
\end{equation}

By using the BPS equations one notices that the energy density (\ref{ED_BPS1})
can also be expressed as%
\begin{equation}
\mathcal{E}_{BPS}=\left(  1-s\right)  B^{2}+2v^{2}\left(  \frac{ag}{r}\right)
^{2}, \label{ED_BPS2}%
\end{equation}
which is a positive-definite expression for $s<1$. \

A first observation is that the Lorentz-violating coefficient does not modify
the minimum energy of the system, given by Eq.(\ref{Hmin}). Under BPS
conditions, the magnetic field is the relevant term for describing the profile
of the energy density associated with the minimum solution. It is also
interesting to note that the BPS equations (\ref{BPS2}) and (\ref{BPS1}) have the
same structure as the BPS equations\textbf{\ }describing\textbf{\ }the AMH
vortex. The difference consists in the presence of the Lorentz-violating
parameter, $s,$ in the second equation given by (\ref{BPS1}), while the first
one remains unchanged. As observed below, the LV parameter acts as
an element able to control both the radial extension and the amplitude of the
defect. The second BPS equation (\ref{BPS1}) can be used to define an
effective electric charge, $e/\sqrt{1-s}$, which holds in the
\textquotedblleft vacuum" of this Lorentz-violating field theory. This
redefinition reveals that this theory can be interpreted as an effective
electrodynamics in a medium pervaded by the Lorentz-breaking tensor background.

In this context, an interesting parallel can to be drawn between the present
model and some effective Maxwell-Higgs Lagrangians in which the Maxwell term
is replaced by $G\left(  \phi\right)  F_{\mu\nu}F^{\mu\nu},$ where $G\left(
\phi\right)  $\ is dubbed the \textquotedblleft dielectric function" because it
introduces a dielectric constant in the equations of motion \cite{Dielectric},
making them similar to the ones that hold in a continuum medium. Under the
vortex ansatz (\ref{Ans2}), the models \cite{Dielectric} provide the following
BPS equation for the magnetic field: $B=\pm\frac{ev^{2}}{G}\left(
1-g^{2}\right)  ,$ which becomes equal to Eq. (\ref{BPS1}) when the
replacement $G\rightarrow(1-s)$ is done. It reveals that the Lorentz-violating model here proposed, whenever subjected to the vortex ansatz\ (\ref{Ans2}), provides vortex solutions in a dielectric medium.

With the purpose of performing the asymptotic and numerical analysis of the
fields in dimensionless form, we introduce the dimensionless variable $t=evr$
and implement the changes\textbf{\ }%
\begin{align}
&  \displaystyle{g\left(  r\right)  \rightarrow\bar{g}\left(  t\right)
,\ \text{\ }a\left(  r\right)  \rightarrow\bar{a}\left(  t\right)
,}~\nonumber\\[-0.2cm]
& \label{changes}\\
&  \displaystyle{B\left(  r\right)  \rightarrow ev^{2}\bar{B}\left(  t\right)
,~~\mathcal{E}_{BPS}\rightarrow v^{2}\bar{\mathcal{E}}_{BPS}.}\nonumber
\end{align}
The BPS equations written in a dimensionless form are%
\begin{align}
\bar{g}^{\prime}  &  =\pm\frac{\bar{a}\bar{g}}{t},\label{BPS3a}\\[0.08in]
-\frac{\bar{a}^{\prime}}{t}  &  =\bar{B}=\pm\frac{1}{1-s}\left(  1-\bar{g}%
^{2}\right)  . \label{BPS3b}%
\end{align}
Notice that Eq. (\ref{BPS3b}), with asymptotic conditions (\ref{bc1}), determines
the magnetic field magnitude at the origin,%
\begin{equation}
\bar{B}(0)=\frac{1}{1-s}. \label{Mag5}%
\end{equation}

\subsection{Asymptotic behavior of the BPS vortex}

Before\ computing the numerical solutions of the BPS equations, we analyze the
asymptotic behavior of the vortex solutions. First, we study the behavior when
$t\rightarrow0$ and solve the BPS equations (\ref{BPS3a}) and (\ref{BPS3b}) by using
a power-series method, achieving
\begin{align}
\bar{g}\left(  t\right)   &  =Gt^{\left\vert n\right\vert }-\frac{G^{2}%
}{4\left(  1-s\right)  }t^{\left\vert n\right\vert +2}+\mathcal{\ldots
},\label{series1}\\[0.3cm]
\bar{a}\left(  t\right)   &  =n-\frac{\ t^{2}}{2\left(  1-s\right)  }%
+\frac{G^{2}~t^{2\left\vert n\right\vert +2}}{2\left(  1-s\right)  (\left\vert
n\right\vert +1)}+\mathcal{\ldots}. \label{series2}%
\end{align}
The specific value of $G$ cannot be determined by the behavior of the fields
around the origin, but it can be fixed by requiring an adequate asymptotic behavior at infinity. A similar situation appears in Ref. \cite{CSV}.

For $t\rightarrow+\infty$, it holds that $\bar{g}=1-\delta g$ and
$\bar{a}=\delta a$, with $\delta g$ and $\delta a$ being small correction
terms. After substituting such forms in (\ref{BPS3a}) and (\ref{BPS3b}), we
obtain the following set of linearized differential equations for $\delta g$
and $\delta a$:%
\begin{equation}
\left(  \delta g\right)  ^{\prime}=-\frac{\delta a}{t},~\ \frac{\left(  \delta
a\right)  ^{\prime}}{t}=-\frac{2\left(  \delta g\right)  }{1-s},
\end{equation}
whose solutions satisfying the appropriate behavior at infinity are%
\begin{align}
\delta g  &  =\left(  \gamma_{s}\right)  ^{-1}K_{0}\left(  \gamma_{s}t\right)
\sim\left(  m_{s}r\right)  ^{-1/2}\exp\left(  -m_{s}r\right)  ,\\[0.35cm]
\delta a  &  =tK_{1}\left(  \gamma_{s}t\right)  \sim\left(  m_{s}r\right)
^{1/2}\exp\left(  -m_{s}r\right)  \text{ ,}%
\end{align}
where $t=evr$, with $\gamma_{s}=\sqrt{2/(1-s)}$. Here, $m_{s}$ is
the mass\ of the bosonic fields, given by%
\begin{equation}
m_{H}=m_{A}=m_{s}=ev\sqrt{\frac{2}{1-s}}\text{.} \label{m}%
\end{equation}
In particular, these asymptotic solutions clearly show how the
Lorentz-violating parameter controls the distance over which the bosons
propagate: the mass increases with $s$. The heavier the bosons are, the
shorter the range of the interaction mediated, and vice versa. Note that
the effective charge (and boson mass) increases while $s$ varies from $0$ to
$1.$ Thus, the Lorentz-violating medium affects the distance over which the
bosons propagate (i.e., the penetration length). In the limit $s\rightarrow1$,
the effective charge and the boson mass diverge, defining an extremely
short-ranged theory. In this limit, the vortex core length tends to zero.
Obviously, there is a correspondence between the interaction range and the
spatial extension of the defect, to be confirmed by analyzing the vortex
profiles. Therefore, the asymptotic analysis of the BPS equations show that
their solutions satisfy the vortex boundary conditions in (\ref{bc1}) and (\ref{bc2}).

\subsection{Numerical solutions for a BPS vortex}

\ Now, we investigate the profiles of the Lorentz-violating BPS solutions
using numerical procedures to solve the differential equations (\ref{BPS3a})
and (\ref{BPS3b}). In particular, we comment on the main aspects in which they differ
from the usual Maxwell-Higgs vortex solutions.

\begin{figure}[H]
\begin{center}
\scalebox{0.9}[0.9]{\includegraphics[width=8cm,height=7cm]{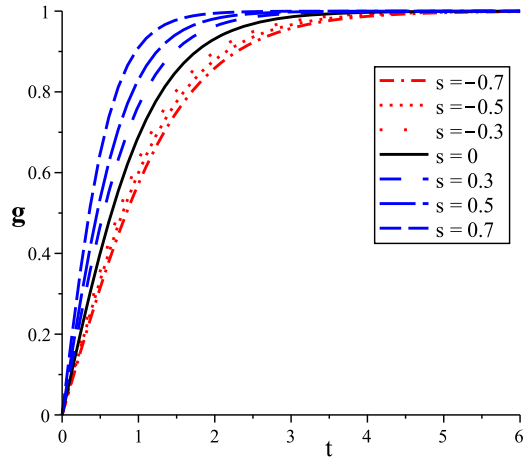}
}
\end{center}
\par
\vspace{-0.5cm}\caption{Scalar field $\bar{g}(t)$ (Solid black line, $s=0$, is
the BPS solution for the Maxwell-Higgs model).}%
\label{S_BPS}%
\end{figure}

\begin{figure}[H]
\begin{center}
\scalebox{0.9}[0.9]{\includegraphics[width=8cm,height=7cm]{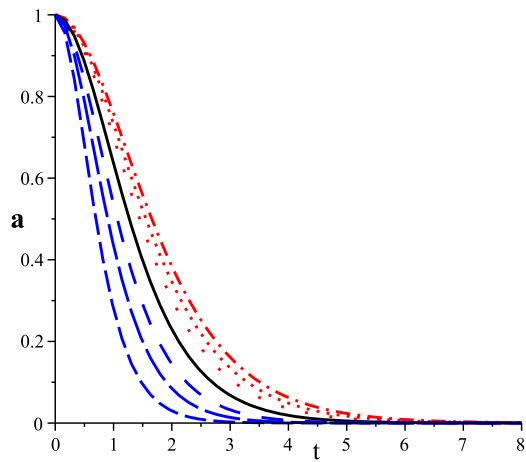}
}
\end{center}
\par
\vspace{-0.5cm}\caption{Vector potential $\bar{a}(t)$.}%
\label{A_BPS}%
\end{figure}

\begin{figure}[H]
\begin{center}
\scalebox{0.9}[0.9]{\includegraphics[width=8cm,height=7cm]{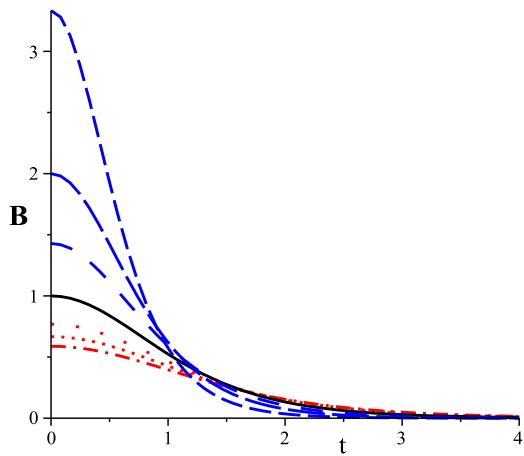}
}
\end{center}
\par
\vspace{-0.5cm}\caption{Magnetic field $\bar{B}(t)$.}%
\label{B_BPS}%
\end{figure}

\begin{figure}[H]
\begin{center}
\scalebox{0.9}[0.9]{\includegraphics[width=8cm,height=7cm]{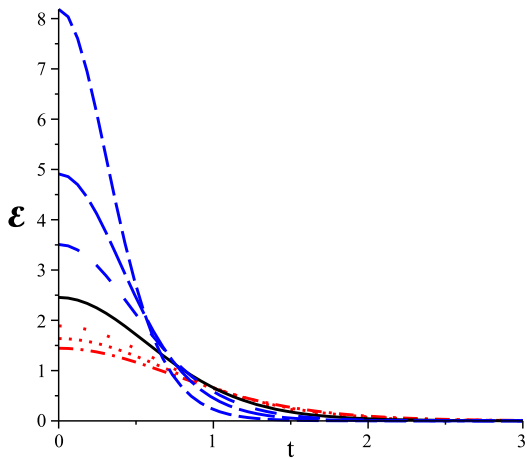}
}
\end{center}
\par
\vspace{-0.5cm}\caption{Energy density $\bar{\mathcal{E}}(t)$.}%
\label{Energy_BPS}%
\end{figure}

In Figs. \ref{S_BPS}--\ref{Energy_BPS}, we present some profiles (for the
winding number $n=1$) for the Higgs field, gauge field, magnetic field and
energy density of the uncharged BPS vortex. This set of graphs reveals the
role played by the Lorentz-violating coefficient, $s$, on the BPS vortex
solutions. In all of them the value $s=0$ reproduces the profile of the vortex
solution of the Maxwell-Higgs model \cite{ANO} which is depicted by a solid
black line. Also, all of the legends are given in Fig. \ref{S_BPS}.

Figures \ref{S_BPS} and \ref{A_BPS} depict the numerical results obtained for
the Higgs field and vector potential, showing that the profiles are drawn
around the ones corresponding to the Maxwell-Higgs model. These profiles
become wider for $s<0$, saturating more smoothly as $s$ becomes more negative.
Otherwise, for an increasing parameter in the range $0<s<1$, the profiles
continuously shrink reaching the smallest thickness for $s\rightarrow1$.

Figure \ref{B_BPS} depicts the magnetic field behavior. The profiles are lumps
centered at the origin whose amplitudes are proportional to $(1-s)^{-1}$;
hence, for $s\rightarrow1$\ higher amplitudes and narrower profiles are
obtained. For $s<0$\ and increasing values of $|s|$\ the magnetic field
profile becomes wider and wider, while its intensity continuously diminishes.

Figure \ref{Energy_BPS} shows the energy density profiles which\textbf{\ }are
very similar to the magnetic field ones, but are more localized and possess
greater amplitudes. This is an expected result: once the extension of the
defect is reduced, its amplitude should increase in order to keep the total
energy constant.

Both profiles belonging to the magnetic field and the energy density are
useful to estimate the extension of the defect in the radial dimension. We
thus note that the defect shrinks (becoming a compactlike structure) while the
parameter increases inside the range $0<s<1$. It indicates that Lorentz
violation works as a factor able to reduce the extension of the defect
profiles. Such a reduction occurs simultaneously to the diminishment of the
interaction range, revealing the consistency of this description. In the limit
$s\rightarrow1$, the theory provides a nearly null core vortex (compatible
with a nearly null range).

\section{Uncharged compactlike BPS vortices with fractional magnetic flux in a
Lorentz-violating AMH model}

In this section, we investigate uncharged vortices in a broader LV
environment, keeping non-null Lorentz-violating terms in both the gauge and
Higgs sectors of the Lagrangian (\ref{L2}).\textbf{\ }In order to restrain our
study to uncharged vortices ($A_{0}=0)$ , we should require
\begin{equation}
\kappa_{0i}=0,\text{ }\left(  k_{\phi\phi}\right)  _{0i}=0,\text{ }\left(
k_{\phi F}\right)  _{0j}=0, \label{Unch}%
\end{equation}
for which Lagrangian (\ref{L2}) is reduced to%
\begin{align}
\mathcal{L}_{1+3}  &  =\mathcal{-}\frac{1}{4}F_{ij}F^{ij}-\frac{1}{2}%
\kappa^{ij}F_{i\sigma}F_{j}^{\text{ \ }\sigma}\nonumber\\[0.3cm]
&  +\left\vert \mathcal{D}_{i}\phi\right\vert ^{2}+\left(  k_{\phi\phi
}\right)  ^{ij}\left(  \mathcal{D}_{i}\phi\right)  ^{\ast}\left(
\mathcal{D}_{j}\phi\right) \\[0.3cm]
&  -\frac{1}{2}\left(  k_{\phi F}\right)  ^{ij}F_{ij}\left\vert \phi
\right\vert ^{2}-U\left(  \left\vert \phi\right\vert \right)  .\nonumber
\end{align}

Under the temporal gauge the Gauss's law is trivially solved,\ and Eqs.
(\ref{NAmpere1}) and (\ref{NHiggs2}) are reduced to the form
\begin{align}
&  \hspace{0cm}\left(  \epsilon_{jbc}-\epsilon_{jac\ }\kappa_{ab}-\kappa
_{ja}\epsilon_{abc}\right)  \!\partial_{b}B_{c}+\left(  k_{\phi F}\right)
_{ij}\partial_{j}\left[  \phi^{\ast}\phi\right] \nonumber\\[0.3cm]
&  \hspace{3cm} =eJ_{i}-e\left(  k_{\phi\phi}\right)  _{ij}J_{j},
\label{ampere2}%
\end{align}%
\begin{equation}
\left[  \mathcal{\delta}_{ij}-\left(  k_{\phi\phi}\right)  _{ij}\right]
\mathcal{D}_{i}\mathcal{D}_{j}\phi-\left(  k_{\phi F}\right)  ^{ij}F_{ij}\phi-\frac{\partial
U}{\partial\phi^{\ast}}=0. \label{higgs2}%
\end{equation}

In order to search for stable vortex configurations, we work in cylindrical
coordinates $\left(  r,\theta,z\right)  ,$\ \ implementing a modified vortex
ansatz
\begin{equation}
\phi=\frac{v}{\Gamma^{1/2}}g\left(  r\right)  e^{i n\theta/\Lambda},~\ A_{\theta
}=-\frac{a\left(  r\right)  -n/\Lambda}{er}, \label{NAnsatz}%
\end{equation}
whose form yields regular behavior for the system at $r=0$, whenever the
following boundary conditions are satisfied\
\begin{equation}
a\left(  0\right)  =\frac{n}{\Lambda},~\ \ g\left(  0\right)  =0. \label{bc3}%
\end{equation}
In Eq. (\ref{NAnsatz}), in the absence of LV effects, $v$\ is the vacuum
expectation value of the Higgs field and $n$\ the winding number of the
topological solutions. The parameters $\Gamma$\ and $\Lambda$\ include the
contributions of the Lorentz violation to the vacuum expectation of the Higgs
field and the behavior of the field $a$\ at the origin, respectively.\ As it
will be shown later, the asymptotic analysis at origin of the BPS equations
(\ref{NBPS1}) and (\ref{NBPS2}) reveals that the vortex solutions stemming from this
Lorentz-violating model can only be reconciled with the modified ansatz
(\ref{NAnsatz}). Nevertheless, the magnetic field keeps being defined as
Eq.(\ref{Brot}).

By substituting the ansatz (\ref{NAnsatz}) in the Amp\`{e}re's law
(\ref{ampere2}), one achieves the condition%
\begin{equation}
\left(  k_{\phi\phi}\right)  _{r\theta}=0,
\end{equation}
and the differential equation,%
\begin{equation}
\left(  1-s\right)  B^{\prime}+\frac{v^{2}}{\Gamma}\left(  k_{\phi F}\right)
_{r\theta}\left(  g^{2}\right)  ^{\prime}+2e\frac{v^{2}}{\Gamma}\frac{ag^{2}%
}{r}\left[  1-\left(  k_{\phi\phi}\right)  _{\theta\theta}\right]  =0,
\label{NAmpere2}%
\end{equation}
where $s=\kappa_{rr}+\kappa_{\theta\theta}$.

The Higgs equation of motion (\ref{higgs2}) now is
\begin{align}
&  \left[  1-\left(  k_{\phi\phi}\right)  _{rr}\right]  \left(  g^{\prime
\prime}+\frac{g^{\prime}}{r}\right)  -\left[  1-\left(  k_{\phi\phi}\right)
_{\theta\theta}\right]  \frac{a^{2}}{r^{2}}g\nonumber\\[-0.3cm]
& \label{NHiggs3a}\\
&  \hspace{2cm}+\left(  k_{\phi F}\right)  _{r\theta}Bg+\frac{\Gamma}{2v^{2}%
}\frac{\partial U}{\partial g}=0.\nonumber
\end{align}
This equation can only be written as a first-order differential equation if
we assume a generalized self-duality condition, that is,
\begin{equation}
g^{\prime}=\pm\Lambda\frac{ag}{r}, \label{NBPS1}%
\end{equation}
with the parameter $\Lambda$\ conveniently defined as%
\begin{equation}
\Lambda=\sqrt{\frac{1-\left(  k_{\phi\phi}\right)  _{\theta\theta}}{1-\left(
k_{\phi\phi}\right)  _{rr}}}. \label{Nparam3b}%
\end{equation}

The BPS equation (\ref{NBPS1}) also allows us to integrate the Amp\`{e}re's law
(\ref{NAmpere2}), yielding another first-order equation involving $a^{\prime}$
\begin{equation}
-\frac{a^{\prime}}{er}=B=\pm\frac{ev^{2}}{1-s}\left(  1-g^{2}\right)  ,
\label{NBPS2}%
\end{equation}
which is the second BPS equation. It is exactly the same one obtained in the
previous section (see Eq. (\ref{BPS1})). Then, by using the BPS equations in
(\ref{NHiggs3a}) it is possible to compute the BPS fourth-order
self-interacting potential,\
\begin{equation}
U\left(  g\right)  =\frac{e^{2}v^{4}}{2\left(  1-s\right)  }\left(
1-g^{2}\right)  ^{2}, \label{Pot2}%
\end{equation}
that written in terms of the Higgs field gives%
\begin{equation}
U\left(  \left\vert \phi\right\vert \right)  =\frac{\Gamma^{2}e^{2}}{2\left(
1-s\right)  }\left(  \frac{v^{2}}{\Gamma}-\left\vert \phi\right\vert
^{2}\right)  ^{2}, \label{potBPS}%
\end{equation}
with
\begin{align}
\Gamma &  =\eta\pm\frac{\left(  k_{\phi F}\right)  _{r\theta}}{e},\\[0.3cm]
\eta &  =\sqrt{\left[  1-\left(  k_{\phi\phi}\right)  _{\theta\theta}\right]
\left[  1-\left(  k_{\phi\phi}\right)  _{rr}\right]  }.
\end{align}
It reveals that this theory may support two different vacua,%
\begin{equation}
\left\vert \phi\right\vert =\frac{v}{\sqrt{\eta\pm\displaystyle\frac{\left(
k_{\phi F}\right)  _{r\theta}}{e}}}, \label{Vac}%
\end{equation}
induced by the coefficient $\left(  k_{\phi F}\right)  _{r\theta}.$\ We point
out that for each vacuum in\ (\ref{Vac}) there exists\ topological defect
solutions of definite vorticity, once\ the $\left(  +\right)  $\ vacuum
supports defects having only a positive winding number $\left(  n>0\right)
$\ whereas the $\left(  -\right)  $\ vacuum is related to vortices with
negative winding number $\left(  n<0\right)  $. It is known that the
Maxwell-Higgs solutions present the following correspondence: the ones
represented with $a\left(  r\right)  $\ and $g\left(  r\right)  $\ for
$n>0$\ are mapped in solutions with $n<0$\ by doing $a\left(  r\right)
\rightarrow-a\left(  r\right)  $\ and $g\left(  r\right)  \rightarrow g\left(
r\right)  $. This correspondence is broken in a theory endowed with the vacuum
(\ref{Vac}), which could be associated with the breaking of the discrete
symmetry connecting the solutions with $n>0$\ and $n<0$. The physical
soundness of this hypothesis is still to be verified after analysis concerning
energy stability.

Now, adopting the conditions (\ref{Unch}) and setting the temporal gauge
$\left(  A_{0}=0\right)  $, we evaluate the energy density for this stationary
and uncharged system as
\begin{align}
\mathcal{E}  &  =\frac{1}{2}\left(  1-\kappa_{jj}\right)  B^{2}+\left\vert
\mathcal{D}_{j}\phi\right\vert ^{2}-\left(  k_{\phi\phi}\right)  _{ij}\left(
\mathcal{D}_{i}\phi\right)  ^{\ast}\left(  \mathcal{D}_{j}\phi\right)
\nonumber\\[0.2cm]
&  \quad+\left(  k_{\phi F}\right)  _{12}B\left\vert \phi\right\vert
^{2}+U\left(  \left\vert \phi\right\vert \right)  . \label{energy_2}%
\end{align}
After replacing the ansatz (\ref{NAnsatz}) and implementing the BPS procedure,
it is rewritten as
\begin{align}
\mathcal{E}  &  =\frac{1}{2}\left(  1-s\right)  \left[  B\mp\frac{ev^{2}}%
{1-s}\left(  1-g^{2}\right)  \right]  ^{2}\nonumber\\[0.2cm]
&  \quad+\frac{v^{2}}{\Gamma}\left[  1-\left(  k_{\phi\phi}\right)
_{rr}\right]  \left[  \!\frac{{}}{{}}g^{\prime}\mp\Lambda\frac{ag}{r}\right]
^{2}\label{energy_2n}\\[0.2cm]
&  \quad\pm Bev^{2}\mp e\frac{v^{2}}{\Gamma}\eta Bg^{2}\pm2\frac{v^{2}}%
{\Gamma}\eta\frac{ag}{r}g^{\prime}.\nonumber
\end{align}
Requiring energy minimization, the squared brackets must vanish, yielding the
BPS equations (\ref{NBPS2}) and (\ref{NBPS1}), respectively.

Imposing the BPS conditions in Eq. (\ref{energy_2n}), one achieves the BPS
energy density which is simplified to the form
\begin{equation}
\mathcal{E}_{BPS}=\mp v^{2}\frac{a^{\prime}}{r}\pm\eta\frac{v^{2}}{\Gamma
}\frac{\left(  ag^{2}\right)  ^{\prime}}{r}, \label{EBPSX}%
\end{equation}
whose integration,\ considering the following asymptotic conditions,
\begin{equation}
a\left(  \infty\right)  =0,~\ \ g\left(  \infty\right)  =1, \label{bc4}%
\end{equation}
leads to the total BPS energy%
\begin{equation}
E_{BPS}=\frac{v^{2}}{\Lambda}\left\vert n\right\vert . \label{NTotalE}%
\end{equation}
So, we must notice that the total\ BPS\ energy is not affected by the
difference between the two vacua (\ref{Vac}), associated, in principle, with
the spontaneous breaking of some discrete symmetry.

It is very interesting to note that magnetic flux is now a multiple of the
fractional ratio $2\pi/e\Lambda,$\ that is,
\begin{equation}
\Phi_{B}=\int B\left(  r\right)  d^{2}\mathbf{r}=\frac{2\pi}{e\Lambda}n,
\end{equation}
indicating that this model provides fractional vortex solutions, which have
been recently reported in condensed matter literature \cite{PRL}. Note that
the relation between the total energy and magnetic flux, $H_{\min}%
=ev^{2}\left\vert \Phi_{B}\right\vert ,$\ continues to be valid.

By using the BPS\ equations, the energy density (\ref{EBPSX}) can be written
as%
\begin{equation}
\mathcal{E}_{BPS}=\frac{e^{2}v^{4}\eta}{\left(  1-s\right)  \Gamma}\left(
\frac{\Gamma}{\eta}-g^{2}\right)  \left(  1-g^{2}\right)  +2v^{2}\frac
{\eta\Lambda}{\Gamma}\left(  \frac{ag}{r}\right)  ^{2}, \label{ebps_nonpos}%
\end{equation}
which is not a positive-definite expression. The energy density
(\ref{energy_2}) shows explicitly that the parameter $\left(  k_{\phi
F}\right)  _{r\theta}$\ is responsible for this energy instability. Hence,
one must require $\left(  k_{\phi F}\right)  _{r\theta}=0$\ for assuring
energy stability. Under such condition\ we have $\Gamma=\eta$, and Eq.
(\ref{ebps_nonpos}) provides a positive-definite BPS energy density\
\begin{equation}
\mathcal{E}_{BPS}=\left(  1-s\right)  B^{2}+2v^{2}\Lambda\left(  \frac{ag}%
{r}\right)  ^{2}, \label{ddee}%
\end{equation}
whenever $s<1$\ and\ $\left(  k_{\phi\phi}\right)  _{rr},\left(  k_{\phi
F}\right)  _{\theta\theta}<1$. \ Similarly, with $\left(  k_{\phi F}\right)
_{r\theta}=0$, the truly BPS self-interacting potential becomes
\begin{equation}
U=\frac{e^{2}\eta^{2}}{2\left(  1-s\right)  }\left(  \frac{v^{2}}{\eta
}-\left\vert \phi\right\vert ^{2}\right)  ^{2}, \label{UHg}%
\end{equation}
providing a unique modified vacuum,
\begin{equation}
\left\vert \phi\right\vert \mathbf{=}\frac{v}{\sqrt{\eta}}\mathbf{,}
\label{VHg}%
\end{equation}
that supports solutions with two vorticities, as is usual. So, we highlight
that the consistent uncharged vortex solutions of this model are the ones
ruled by Eqs. (\ref{NBPS1}) and (\ref{NBPS2}), with self-interacting potential
(\ref{UHg}), and the modified vacuum expectation value (\ref{VHg}). Note that
the LV Higgs coefficients modify the usual self-duality condition (\ref{BPS2})
and the vacuum of the Maxwell-Higgs model.

One should still discuss the reason that requires the modified vortex ansatz
(\ref{NAnsatz}). First, note that if we had supposed the usual ansatz
(\ref{Ans2}), one would have achieved the same BPS equations given by
(\ref{NBPS1}) and (\ref{NBPS2}). On the other hand, performing the series expansion
at $r=0$ of the regular functions $g\left(  r\right)  ,a\left(  r\right)  $
fulfilling the BPS\ equations, we obtain
\begin{align}
g\left(  r\right)   &  =Gr^{\left\vert n\right\vert }-\frac{\Lambda G\left(
ev\right)  ^{2}}{4\left(  1-s\right)  }r^{\left\vert n\right\vert
+2}+\mathcal{\ldots},\\[0.3cm]
a\left(  r\right)   &  =\frac{n}{\Lambda}-\frac{\left(  evr\right)  ^{2}%
}{2\left(  1-s\right)  }+\frac{\left(  Gev\right)  ^{2}~r^{2\left\vert
n\right\vert +2}}{2\left(  1-s\right)  \left(  \left\vert n\right\vert
+1\right)  }+\mathcal{\ldots}, \label{Nas}%
\end{align}
where $G$ is a constant. It is easy to notice that the expansion (\ref{Nas})
is incompatible with the usual ansatz (\ref{Ans2}), stating an inconsistency.
In order to avoid it we have adopted the new ansatz (\ref{NAnsatz}), keeping
the field $A_{\theta}$ finite at $r=0.$

We now present the asymptotic behavior at $r=\infty$. By setting%
\begin{equation}
g\left(  r\right)  =1+\delta g,~\ \ \ a\left(  r\right)  =\delta a,
\end{equation}
and after solving the linearized BPS\ equations for $\delta g$ and $\delta a$,
we obtain%
\begin{align}
\delta g  &  =-\frac{G_{1}}{ev}\sqrt{\frac{\Lambda\left(  1-s\right)  }{2}%
}K_{0}\left(  rev\sqrt{\frac{2\Lambda}{1-s}}\right)  ,~\ \ \\[0.2cm]
\delta a  &  =G_{1}~r~K_{1}\left(  rev\sqrt{\frac{2\Lambda}{1-s}}\right)  ,
\end{align}
from which we observe that the mass of the gauge and Higgs fields is
\begin{equation}
m_{A}=m_{H}=ev\sqrt{\frac{2\Lambda}{1-s}}. \label{m2}%
\end{equation}
In this case, notice that there are two LV parameters modifying the mass. For
a fixed $\Lambda$, it holds the behavior described after Eq. (\ref{m}) for
compactlike defects. For a fixed $s,$ we observe that the\ mass increases with
$\Lambda$, in opposition to its dependence with $s.$

In order to facilitate the numerical analysis of the vortex profiles, we first
introduce the dimensionless variable, $t=ev\Lambda^{1/2}r,$ and the following
field redefinitions:
\begin{equation}
g\left(  r\right)  \rightarrow\bar{g}\left(  t\right)  ,~\ \ a\left(
r\right)  \rightarrow\frac{\bar{a}\left(  t\right)  }{\Lambda}; \label{Fresc}%
\end{equation}
with the new variable, the BPS equations (\ref{NBPS1}) and (\ref{NBPS2}) are rewritten as
\begin{equation}
\bar{g}^{\prime}=\pm\frac{\bar{a}\bar{g}}{t},\text{ \ \ }-\frac{\bar
{a}^{\prime}}{t}=\bar{B}=\pm\frac{1-\bar{g}^{2}}{\left(  1-s\right)  },
\end{equation}
assuming the same form as Eqs. (\ref{BPS3a}) and (\ref{BPS3b}). This
shows that under the conditions that provide uncharged vortex configurations
the broader model of Lagrangian (\ref{L2}) also supports compactlike solutions
with controllable size as well.

However, note that the field rescaling (\ref{Fresc}) into the full Lagrangian
(\ref{L2}) does not yield the model of Lagrangian (\ref{L1}) at all. \ It is
important to note that a simple field rescaling in the Lagrangian (\ref{L1})
does not lead to the fractional vortices engendered by Lagrangian (\ref{L2}).
In this sense, note that the rescaling$\ \bar{a}\rightarrow\bar{a}/\Lambda
$\ whenever applied into\ Eq. (\ref{series2})\ leads to
\begin{equation}
\bar{a}\left(  t\right)  =\frac{n}{\Lambda}-\frac{\ t^{2}}{2\left(
1-s\right)  \Lambda}+\frac{G^{2}~t^{2\left\vert n\right\vert +2}}{2\left(
1-s\right)  (\left\vert n\right\vert +1)\Lambda}+\mathcal{\ldots}.
\label{series5}%
\end{equation}
This equation is consistent with the ansatz (\ref{NAnsatz}), indicating
fractional magnetic flux. Nevertheless, it does not represent the original
magnetic field for the uncharged vortex stemming from Lagrangian (\ref{L1}).
Indeed, while this series leads to $B(0)=\left[  \left(  1-s\right)
\Lambda\right]  ^{-1}$, the correct result is the one of Eq. (\ref{Mag5}).
This means that the physical equivalence between the models belonging to the
Lagrangians (\ref{L2}) and (\ref{L1}) could be established by performing
adequate coordinates and field transformations, as discussed below using polar
coordinates and by means of a general coordinate system in the Appendix.

Once the BPS equations are equal, and the asymptotic conditions are
qualitatively similar, one asserts that the vortex profiles stemming from Eqs.
(\ref{NBPS1}) and (\ref{NBPS2}) present the same behavior of the ones depicted in
Figs. \ref{S_BPS}, \ref{A_BPS}, \ref{B_BPS}, \ref{Energy_BPS}. In this way, a
numerical evaluation of the new corresponding profiles becomes unnecessary.

There exist\ some maps connecting the models of Lagrangians (\ref{L2}) and
(\ref{L1}).\textbf{ }For example, in polar coordinates such mapping is easily
observed: it involves a simultaneous rescaling of the radial coordinate
and\ the gauge field. By writing the energy density (\ref{energy_2}) with
$\left(  k_{\phi F}\right)  _{12}=0$\ in polar coordinates, we have%
\begin{align}
E  &  =2\pi%
{\displaystyle\int}
dr~r\left[  \frac{1}{2}\left(  1-s\right)  B^{2}+\frac{v^{2}}{\Lambda}\left(
g^{\prime}\right)  ^{2}+v^{2}\Lambda\frac{\left(  ag\right)  ^{2}}{r^{2}%
}\right. \\
&  ~\ \ \ \ \ \left.  ~\ \ \ \ \ \ \ \ \ \ \ \ +\frac{v^{4}e^{2}}{2\left(
1-s\right)  }\left(  1-g^{2}\right)  ^{2}\right]  .\nonumber
\end{align}
Performing the rescaling,%
\begin{equation}
r=\frac{\bar{r}}{\Lambda^{1/2}},\ \ a\left(  r\right)  \rightarrow
\frac{a\left(  \bar{r}\right)  }{\Lambda},~\ \ g\left(  r\right)  \rightarrow
g\left(  \bar{r}\right)  ,~ \label{Transf2}%
\end{equation}
we obtain%
\begin{align}
E  &  =\frac{2\pi}{\Lambda}%
{\displaystyle\int}
d\bar{r}~\bar{r}\left\{  \frac{1}{2}\left(  1-s\right)  \left(  \frac
{a^{\prime}}{e\bar{r}}\right)  ^{2}+v^{2}\left[  \left(  g^{\prime}\right)
^{2}+\frac{\left(  ag\right)  ^{2}}{\bar{r}^{2}}\right]  \right. \nonumber\\
&  ~\ \ \ \ \left.  \ ~\ \ \ \ \ \ \ \ \ \ \ \ +\frac{v^{4}e^{2}}{2\left(
1-s\right)  }\left(  1-g^{2}\right)  ^{2}\right\}  .
\end{align}
The term in bracket is the energy density (\ref{H}), expressed in polar
coordinates,\ for the vortices of the simpler model studied in Sec. III. In
Cartesian coordinates this mapping would be more involved, including a
rotation followed by a coordinate rescaling in order to transform the Higgs
sector of the Lagrangian (\ref{L2}) into the one of (\ref{L1}). Obviously, such
transformations would also affect the gauge sector by adding LV Higgs
contributions to the already existent LV gauge parameters. At this point, it
would still be necessary to perform a gauge field rescaling in order to
transform completely (\ref{L2}) into (\ref{L1}). This set of transformations is
easily performed in polar coordinates, as shown in (\ref{Transf2}).

\section{Conclusions}

In this work, we have investigated the existence of stable\ uncharged BPS
vortex configurations in the framework of an Abelian Maxwell-Higgs model
supplemented with CPT-even and Lorentz-violating terms belonging to the gauge
and Higgs sectors of the standard model extension. Our study has accounted two
situations: the first one considered LV terms only in the gauge sector while
the second case\ regarded the full model. In both cases we have\ restrained
the investigation to uncharged topological solutions. In the first case, we
have found the equations of motion and implemented the usual ansatz for static
and rotationally symmetric vortex solutions. By applying the Bogomol'nyi
method to the energy density, the first-order BPS equations were achieved,
exhibiting a similar structure to the ones of the Abelian Maxwell-Higgs model
which supports the ANO vortices. A numerical procedure was used to unveil the
profiles of the BPS vortex solutions. Although the governing equations present
the same structure, the LIV parameter, $s$, appears as a key element that
allows us to control the thickness (radial extent) of the defect while spanning the available range $\left(  s<1\right)  $: the larger is the value of the LIV
parameter, $s$; the narrower is the profile of the vortex, and vice versa. The
intensity of the magnetic field increases with $s$, tending to a maximal value
when $s\rightarrow1,$ a limit in which the length of the defect and the
interaction range tends to zero. Our results offer the possibility of
controlling the extension of the defect without modifying the kinetic sector
or the Higgs potential of the model. In this theoretical framework, Lorentz
violation (for $0<s<1$) plays a role similar to some nonlinear kinetic terms
usual in \emph{k}-field theories, which yield compactlike defects
\cite{Hora1,Compact1,kfield,Hora2,Hora3}. Despite the profile shrinking
observed here, analogous to the one verified in \emph{k}-field compactlike defects
(see Ref. \cite{Hora2}), one should point out some advantages of the
Lorentz-violating compactlike solutions: the BPS character, the preservation
of the usual kinetic sector of the Maxwell-Higgs theory, and the direct
correspondence between the defect thickness and the range of the interaction.
Moreover, this result opens a new window: the chance of employing models with
Lorentz violation as an effective theory to address some situations wherein
\emph{k}-field models have been applied (see
Refs.\cite{Cosmology,DM,Tachyon,GC,kfield}), with special attention to
topological defects. More specifically, the possibility of controlling the
interaction range and the size of the defect may allow interesting
applications to the investigation of vortex configurations in some condensed
matter frameworks, where the penetration length depends on the properties of system.

One should remark that the vortex solutions provided by Eqs. (\ref{BPS2})
and (\ref{BPS1}) do not exhibit space anisotropy, although the LV parity-even
coefficients usually yield anisotropic stationary solutions (see
Ref.\cite{Paulo1}). This indicates that the vortex ansatz selects only the
parity-even solutions not endowed with space anisotropy. We still remember
that the vortex configurations obtained in this Lorentz-violating model are
somehow equivalent to the ones yielded by the effective Maxwell-Higgs
electrodynamics of Ref. \cite{Dielectric}, which describes vortex
configurations in a continuum medium. Note that in this situation it does not
make sense to consider vacuum upper bounds on the Lorentz-violating
coefficients, once the role of the Lorentz-violating coefficient is to state
the presence of the dielectric medium. This interpretation turns sensible the
profile analysis performed in this work for the range $-1<s<1$, which obviously
involves magnitudes much higher than the upper bounds usually stated, for
example, in an electrodynamics in the vacuum.

Lastly, we have addressed the case where the Lorentz violation is present in
the Higgs and gauge sectors simultaneously. The equations of motion were
determined, being possibly compatible with BPS equations defined in the context
of a theory with two different vacua, each one supporting a definite vorticity
solution. The energy density was carried out, revealing that only one of the
Higgs LV coefficients yields solutions with energy stability, so that we have
adopted $\left(  k_{\phi F}\right)  _{r\theta}=0.$\ The remaining theory
provides stable uncharged vortices (with both vorticities) in a unique
vacuum.\ It has implied BPS equations consisting of a modified duality
relation and compatible with a new rotationally symmetric vortex ansatz. After
an appropriated coordinate and field rescaling, such equations recover the BPS
equations of the first case (with Lorentz violation only in the gauge sector),
leading to vortex solutions with similar profiles but different magnetic flux.
In this case, however, the defect profiles are controlled by two LV
parameters, $\Lambda$ and $s,$ that play opposite roles on the spatial
extension and the range of the interaction. For a fixed $\Lambda,$ we recover
the same phenomenology described for the first case. For a fixed $s,$ the size
of the vortex should diminish with an increasing $\Lambda.$ Hence, we notice
that $\Lambda$\ and $s$\ act as competing parameters, providing a larger
control on the solution. \ The crucial difference between the first and the
second model is entailed with the new vortex ansatz\textbf{\ }required to
imply consistent solutions at origin. Such difference engenders an interesting
novelty: vortices with fractional magnetic flux, which is a feature of
interest in condensed matter systems, as, for example, in recent models for
superconductivity \cite{PRL} where the vortex solutions have a fractional
structure. Finally, we highlight that the fractional BPS solutions are
explicitly defined in the context of a modified vacuum theory, in accordance
with Eq.(\ref{VHg}). The point is that in the presence of the LV-Higgs parameters,
the self-interacting fourth-order potential must be given as in Eq.
(\ref{UHg}) for ensuring the existence of BPS solutions. \textbf{\ }In the
Appendix, we discuss the equivalence between the models of Lagrangians
(\ref{L2}) and (\ref{L1}). Such an equivalence leads to the conclusion that
the simpler model of Lagrangian (\ref{L1}) should also possess fractional
solutions, in principle uncovered, and only revealed by suitable coordinate transformations.

New developments in this Lorentz-violating environment are now under way,
mainly in connection with the search for charged vortex configurations, in the
absence of the Chern-Simons term, when the Higgs sector is supplemented with a
richer self-interacting potential \cite{Carlisson2}.\textbf{\ }Finally, an
interesting investigation would be to verify the existence of\ topological
defects in the non-Abelian sector of the SME.

\section{Appendix}

In this Appendix we comment on the existence of a coordinate transformation
stating the equivalence of the models of Lagrangians (\ref{L2}) and (\ref{L1})
at first order in the Lorentz-violating parameters. This coordinate
transformation can be generally written as
\begin{align}
x^{\mu}  &  =\left(  M^{-1}\right)  ^{\mu}{}_{\nu}x^{\prime\nu},~\ \ \partial
_{\mu}=M^{\nu}{}_{\mu}\partial_{\nu}^{\prime},\nonumber\label{Ctrans1}\\
& \\[-0.2cm]
A_{\mu}  &  =M^{\nu}{}_{\mu}A_{\nu}^{\prime},~\ \mathcal{D}_{\mu}\phi=M^{\nu
}{}_{\mu}\mathcal{D}_{\nu}^{\prime}\phi.\nonumber
\end{align}
We can show that this transformation yields%
\begin{align}
\mathcal{L}_{\phi}  &  =\left\vert \mathcal{D}_{\mu}\phi\right\vert
^{2}+\left(  k_{\phi\phi}\right)  ^{\mu\nu}\left(  \mathcal{D}_{\mu}%
\phi\right)  ^{\ast}\left(  \mathcal{D}_{\nu}\phi\right) \nonumber\\[0.3cm]
&  =\left[  g^{\mu\nu}+\left(  k_{\phi\phi}\right)  ^{\mu\nu}\right]  \left(
\mathcal{D}_{\mu}\phi\right)  ^{\ast}\left(  \mathcal{D}_{\nu}\phi\right)
\nonumber\\[0.3cm]
&  =g^{\prime\alpha\beta}\left(  \mathcal{D}_{\alpha}^{\prime}\phi\right)
^{\ast}\left(  \mathcal{D}_{\beta}^{\prime}\phi\right)  ,
\end{align}
where $g^{\prime\alpha\beta}$ is a new metric tensor related to $g^{\mu\nu}$
via the equation%
\begin{equation}
\left[  g^{\mu\nu}+\left(  k_{\phi\phi}\right)  ^{\mu\nu}\right]  M^{\alpha}%
{}_{\mu}M^{\beta}{}_{\nu}=g^{\prime\alpha\beta}. \label{link1}%
\end{equation}
At the first order, $M^{\alpha}{}_{\mu}$\ can be written as%
\begin{equation}
M^{\alpha}{}_{\mu}=\delta^{\alpha}{}_{\mu}+m^{\alpha}{}_{\mu}, \label{Mtr1}%
\end{equation}
which, when replaced in Eq.(\ref{link1}) leads to%
\begin{equation}
g^{\alpha\beta}+m^{\beta\alpha}+m^{\alpha\beta}+\left(  k_{\phi\phi}\right)
^{\alpha\beta}=g^{\prime\alpha\beta}. \label{Metric1}%
\end{equation}
At this point, we can organize the possible maps into two possibilities.

\subsection{First case}

A preliminary case is addressed when one requires
\begin{equation}
m^{\beta\alpha}=m^{\alpha\beta}=-\frac{1}{2}\left(  k_{\phi\phi}\right)
^{\alpha\beta},
\end{equation}
so that Eq. (\ref{Mtr1}) reads as
\begin{equation}
M^{\alpha}{}_{\mu}=\delta^{\alpha}{}_{\mu}-\frac{1}{2}\left(  k_{\phi\phi
}\right)  ^{\alpha}{}_{\mu}, \label{Mtr2}%
\end{equation}
and the metric remains unaffected, $g^{\alpha\beta}=g^{\prime\alpha\beta}.$
Within this context, one explicitly evaluates%
\begin{align}
\mathcal{L}_{F}  &  =\mathcal{-}\frac{1}{4}F_{\mu\nu}F^{\mu\nu}-\frac{1}%
{2}\kappa^{\rho\alpha}F_{\rho\sigma}F_{\alpha}^{\text{ \ }\sigma
}\nonumber\\[0.3cm]
&  =\mathcal{-}\frac{1}{4}F_{\alpha\beta}^{\prime}F^{\prime\alpha\beta}%
-\frac{1}{2}\kappa^{\prime\alpha\delta}F_{\alpha\beta}^{\prime}F_{\delta
}^{\prime}~^{\beta},
\end{align}
where we have defined%
\begin{equation}
\kappa^{\prime\alpha\delta}=\kappa^{\alpha\delta}-\left(  k_{\phi\phi}\right)
^{\alpha\delta}. \label{klinha1}%
\end{equation}

Therefore, under the coordinate transformation (\ref{Mtr2}), and at first
order in the LV parameters, the full Lagrangian,
\begin{align}
\mathcal{L}  &  =\mathcal{-}\frac{1}{4}F_{\mu\nu}F^{\mu\nu}-\frac{1}{2}%
\kappa^{\rho\alpha}F_{\rho\sigma}F_{\alpha}^{\text{ \ }\sigma}+\left\vert
\mathcal{D}_{\mu}\phi\right\vert ^{2}\label{FullLag1}\\[0.12in]
&  +\left(  k_{\phi\phi}\right)  ^{\mu\nu}\left(  \mathcal{D}_{\mu}%
\phi\right)  ^{\ast}\left(  \mathcal{D}_{\nu}\phi\right)  -U\left(  \left\vert
\phi\right\vert \right)  ,\nonumber
\end{align}
becomes equivalent to%
\begin{equation}
\mathcal{L}=\mathcal{-}\frac{1}{4}F_{\mu\nu}^{\prime}F^{\prime\mu\nu}-\frac
{1}{2}\kappa^{\prime\rho\alpha}F_{\rho\sigma}^{\prime}F_{\alpha}^{\prime\text{
\ }\sigma}+\left\vert \mathcal{D}_{\mu}^{\prime}\phi\right\vert ^{2}-U\left(
\left\vert \phi\right\vert \right)  , \label{NewLag1}%
\end{equation}
revealing that the Lorentz-violating terms were moved from the scalar to the
gauge sector.

\subsection{Second case}

Another possibility consists in taking
\begin{equation}
m^{\alpha\beta}=-m^{\beta\alpha},
\end{equation}
so that $M^{\alpha}{}_{\mu}$ is an orthogonal matrix at first order. The
metric relation \textbf{(}\ref{Metric1}\textbf{) } is now%
\begin{equation}
g^{\prime\alpha\beta}=g^{\alpha\beta}+\left(  k_{\phi\phi}\right)
^{\alpha\beta},
\end{equation}
representing a nondiagonal matrix. At first order,%
\begin{align}
M^{\alpha}{}_{\mu}M^{\beta}{}_{\nu}  &  =\delta^{\alpha}{}_{\mu}\delta^{\beta
}{}_{\nu}+\delta^{\alpha}{}_{\mu}m^{\beta}{}_{\nu}+m^{\alpha}{}_{\mu}%
\delta^{\beta}{}_{\nu},\nonumber\\
& \label{FO3b}\\
g^{\mu\nu}M^{\alpha}{}_{\mu}M^{\beta}{}_{\nu}  &  =g^{\alpha\beta}~,~\left(
k_{\phi\phi}\right)  ^{\mu\nu}M^{\alpha}{}_{\mu}M^{\beta}{}_{\nu}=\left(
k_{\phi\phi}\right)  ^{\alpha\beta}.\nonumber
\end{align}

Using the first-order relations (\ref{FO3b}), after some algebra we can
explicitly show that
\begin{align}
\mathcal{L}_{F}  &  =\mathcal{-}\frac{1}{4}F_{\mu\nu}F^{\mu\nu}-\frac{1}%
{2}\kappa^{\rho\alpha}F_{\rho\sigma}F_{\alpha}^{\text{ \ }\sigma}\nonumber\\
&  =\mathcal{-}\frac{1}{4}F_{\mu\nu}^{\prime}F^{\prime\mu\nu}-\frac{1}%
{2}\kappa^{\prime\rho\alpha}F_{\rho\sigma}^{\prime}F_{\alpha}^{\prime\text{
\ }\sigma},
\end{align}
with the redefinition (\ref{klinha1}). Therefore, at first order in LIV
parameters, under the coordinate transformation (\ref{Ctrans1}), the full
Lagrangian (\ref{FullLag1}) becomes%
\begin{align}
\mathcal{L}^{\prime}  &  =\mathcal{-}\frac{1}{4}F_{\mu\nu}^{\prime}%
F^{\prime\mu\nu}-\frac{1}{2}\kappa^{\prime\rho\alpha}F_{\rho\sigma}^{\prime
}F_{\alpha}^{\prime}{}^{\sigma}\nonumber\\
&  +g^{\prime\mu\nu}\left(  \mathcal{D}_{\mu}^{\prime}\phi\right)  ^{\ast
}\left(  \mathcal{D}_{\nu}^{\prime}\phi\right)  -U\left(  \left\vert
\phi\right\vert \right)  , \label{NewLag2}%
\end{align}
with the observation that the metric $g^{\prime\mu\alpha}$ is nondiagonal. It can
be achieved as a set of transformations (at first order) turning $g^{\prime\mu
\nu}$ diagonal, and stating the equivalence of the models of Lagrangians
(\ref{FullLag1}) and (\ref{NewLag2}).

We have thus stated the equivalence between the models (\ref{FullLag1}) and
(\ref{NewLag1}) at first order in the Lorentz-violating parameters.
Notwithstanding, a more involved transformation may be found assuring the full
equivalence at any order. This fact is related to physical observability of
the LV parameters in the scalar and gauge sectors of the model. For a more
complete discussion about this point, see Sec. II, part C, of Ref.
\cite{Gravity2}. The physical equivalence of the models of Lagrangians
(\ref{L2}) and (\ref{L1}) [with $\left(  k_{\phi F}\right)  ^{\mu\nu}=0$]
leads to the conclusion that the fractional vortex configurations are explicit
solutions of the model of Lagrangian (\ref{L2}) and hidden solutions of the
Lagrangian (\ref{L1}).

\begin{acknowledgments}
The authors are grateful to CNPq, CAPES and FAPEMA (Brazilian research
agencies) for invaluable financial support. These authors also acknowledge the
Instituto de F\'{\i}sica Te\'{o}rica (UNESP-S\~{a}o Paulo State University)
for the kind hospitality during the realization of this work.
\end{acknowledgments}


\begin{thebibliography}{99}                                                                                               %


\bibitem {ANO}A. Abrikosov, Sov. Phys. JETP \textbf{32}, 1442 (1957); H.
Nielsen and P. Olesen, Nucl. Phys.\textbf{B}61, 45 (1973).

\bibitem {CS}S. Deser, R. Jackiw, and S. Templeton, Ann. Phys. (NY)
\textbf{140}, 372 (1982); G.V. Dunne, arXiv:hep-th/9902115.

\bibitem {CSV}R. Jackiw and E. J. Weinberg, Phys. Rev. Lett. \textbf{ 64},
2234 (1990); R. Jackiw, K. Lee, and E.J. Weinberg, Phys. Rev. 
\textbf{D}42, 3488 (1990); J. Hong, Y. Kim, and P.Y. Pac, Phys. Rev.
Lett.\textbf{\ 64}, 2230 (1990); G.V. Dunne, \emph{Self-Dual Chern-Simons Theories}
(Springer, Heidelberg, 1995).

\bibitem {Ezawa}Z.F. Ezawa, \emph{Quantum Hall Effects},  ( World Scientific, 2000 ).

\bibitem {CSV1}P.K. Ghosh, Phys. Rev.\textbf{\ }D\textbf{\ 49}, 5458 (1994);
T. Lee and H. Min, Phys. Rev.\textbf{\ }D\textbf{\ 50}, 7738 (1994).

\bibitem {CSV2}N. Sakai and D. Tong, J. High Energy Phys. \textbf{03}, 019 (2005); G. S.
Lozano, D. Marques, E. F. Moreno, and F. A. Schaposnik, Phys. Lett. B
\textbf{654}, 27 (2007).

\bibitem {Bolog}S. Bolognesi and S.B. Gudnason, Nucl. Phys. B\textbf{805},
104 (2008).

\bibitem {Hora1}D. Bazeia, E. da Hora, C. dos Santos, and R. Menezes, Phys.
Rev.\textbf{\ }D\textbf{\ 81}, 125014 (2010).

\bibitem {Compact1}C. dos Santos, Phys. Rev.\textbf{\ }D\textbf{\ 82}, 125009 (2010).

\bibitem {Hora1b}D. Bazeia, E. da Hora, C. dos Santos, and R. Menezes, Eur. Phys.
J. C\textbf{\ 71}, 1833 (2011).

\bibitem {HoraTwin}D. Bazeia, E. da Hora, and R. Menezes, Phys.
Rev.\ D\textbf{\ 85}, 045005 (2012).

\bibitem {Cosmology}C. Armendariz-Picon, T. Damour, and V. Mukhanov, Phys. Lett. B
\textbf{458}, 209 (1999).

\bibitem {DM}C. Armendariz-Picon and E. A. Lim, J. Cosmol. Astropart. Phys. \textbf{08}, 007 (2005).

\bibitem {Tachyon}A. Sen, J.High Energy Phys. \textbf{07}, 065 (2002).

\bibitem {GC}N. Arkani-Hamed, H.-C. Cheng, M. A. Luty, and S. Mukohyama, J.High Energy Phys. \textbf{05}, 074 (2004); N. Arkani-Hamed, P. Creminelli, S. Mukohyama, and M.
Zaldarriaga, J. Cosmol. Astropart. Phys. \textbf{04}, 001 (2004); S. Dubovsky, J. Cosmol. Astropart. Phys. \textbf{07}, 009 (2004); D. Krotov, C. Rebbi, V. Rubakov, and V. Zakharov, Phys.Rev. D \textbf{71}, 045014
(2005); A. Anisimov and A. Vikman, J. Cosmol. Astropart. Phys. \textbf{04}, 009 (2005).

\bibitem {kfield}E. Babichev, Phys. Rev. D \textbf{74}, 085004 (2006); Phys.
Rev.\textbf{\ }D\textbf{\ 77}, 065021 (2008).

\bibitem {Hora2}D. Bazeia, E. da Hora, R. Menezes, H. P. de Oliveira, and C.
dos Santos, Phys. Rev.\textbf{\ D 81}, 125016 (2010).

\bibitem {Hora3}C. dos Santos and E. da Hora, Eur. Phys. J. C \textbf{70},
1145 (2010); \ Eur. Phys. J. C \textbf{71}, 1519 (2011); D. Bazeia, E. da
Hora, D. Rubiera-Garcia, Phys. Rev. D \textbf{84}, 125005 (2011).

\bibitem {Compact2}P. Rosenau and J.M. Hyman, Phys. Rev. Lett.\textbf{\ 70},
564 (1993); P. Rosenau and E. Kashdan, Phys. Rev. Lett.\textbf{\ 104}, 034101(2010).

\bibitem {Compact3}C. Adam, P. Klimas, J. S\'{a}nchez-Guill\'{e}n, and A.
Wereszczy\'{n}ski, J. Math. Phys. \textbf{50}, 102303 (2009); C. Adam, N.
Grandi, J. Sanchez-Guillen and A. Wereszczynski, J. Phys. A
\textbf{41}, 212004 (2008); C. Adam, P Klimas, J S\'{a}nchez-Guill\'{e}n and
A. Wereszczynski, J. Phys. A \textbf{42}, 135401 (2009).

\bibitem {Colladay}D. Colladay and V. A. Kostelecky, Phys. Rev. D \textbf{55},
6760 (1997); D. Colladay and V. A. Kostelecky, Phys. Rev.\textit{\ }D
\textbf{58}, 116002 (1998); S. R. Coleman and S. L. Glashow, Phys.
Rev.\textbf{\ }D\textbf{\ 59}, 116008 (1999); S.R. Coleman and S.L. Glashow,
Phys. Rev. D\textbf{\ 59}, 116008 (1999).

\bibitem {Samuel}V. A. Kostelecky and S. Samuel, Phys. Rev. Lett. \textbf{63},
224 (1989); Phys. Rev. Lett. \textbf{66}, 1811 (1991); Phys. Rev.\textit{\ }%
D\textbf{\ 39}, 683 (1989); Phys. Rev.\textit{\ }D\textbf{\ 40}, 1886 (1989);
V. A. Kostelecky and R. Potting, Nucl. Phys. B\textbf{ 359}, 545 (1991);
Phys. Lett. B\textbf{381}, 89 (1996); V. A. Kostelecky and R. Potting, Phys.
Rev. D\textbf{\ 51}, 3923 (1995).

\bibitem {Fermion}B. Altschul, Phys. Rev. D \textbf{70}, 056005 (2004); G. M.
Shore, Nucl. Phys. B\textbf{717}, 86 (2005); \ D. Colladay and V. A.
Kostelecky, Phys. Lett. B \textbf{511}, 209 (2001); O. G. Kharlanov and V. Ch.
Zhukovsky, J. Math. Phys. \textbf{48}, 092302 (2007); R. Lehnert, Phys. Rev. D
\textbf{68}, 085003 (2003); V.A. Kostelecky and C. D. Lane, J. Math. Phys.
\textbf{40}, 6245 (1999); R. Lehnert, J. Math. Phys. \textbf{45}, 3399 (2004);
V. A. Kostelecky and R. Lehnert, Phys. Rev. D\textbf{\ 63 }, 065008
(2001); W. F. Chen and G. Kunstatter, Phys. Rev. D \textbf{62},
105029 (2000); B. Goncalves, Y. N. Obukhov, and I. L. Shapiro, Phys.Rev.D
\textbf{80}, 125034 (2009).

\bibitem {Fermion2}M. Gomes, J. R. Nascimento, A. Yu. Petrov, and A. J. da Silva, Phys. Rev. D \textbf{81}, 045018 (2010); T. Mariz, J. R. Nascimento, A. Yu.
Petrov, Phys. Rev. D \textbf{85}, 125003 (2012); T. Mariz, J. R. Nascimento,
and A.Yu. Petrov, Phys. Rev. D \textbf{85}, 125003 (2012); G. Gazzola, H. G.
Fargnoli, A. P. Baeta Scarpelli, M. Sampaio, and M. C. Nemes, J. Phys. G
\textbf{39}, 035002 (2012); A. P. Baeta Scarpelli, Marcos Sampaio, M. C.
Nemes, and B. Hiller, Eur. Phys. J. C \textbf{56}, 571 (2008); F.A. Brito, L.S.
Grigorio, M.S. Guimaraes, E. Passos, and C. Wotzasek, Phys.Rev. D \textbf{78},
125023 (2008); F.A.Brito, E. Passos, and P.V. Santos, Europhys. Lett. \textbf{95},
51001 (2011).

\bibitem {Fermion3}K. Bakke and H. Belich, J. Phys. G \textbf{39},085001 (2012);
K. Bakke, H. Belich, and E. O. Silva, J. Math. Phys. \textbf{52}, 063505 (2011);
\ J. Phys. G \textbf{39}, 055004 (2012); Ann. Physik (Leipzig)
\textbf{523}, 910 (2011).

\bibitem {Gravity}V. A. Kostelecky, Phys. Rev. D \textbf{69}, 105009 (2004);
V. A. Kostelecky, N. Russell, and J. D. Tasson, Phys. Rev. Lett.
\textbf{100}, 111102 (2008); V. A. Kostelecky and J. D. Tasson, Phys. Rev.
Lett. \textbf{102}, 010402 (2009); Q. G. Bailey and V.A. Kostelecky, Phys.Rev. D
\textbf{74}, 045001 (2006); Q. G. Bailey, Phys.Rev. D \textbf{80}, 044004
(2009); V.A. Kostelecky and R. Potting, Phys.Rev. D \textbf{79}, 065018
(2009); Q. G. Bailey, Phys. Rev. D \textbf{82}, 065012 (2010); V.B. Bezerra,
C.N. Ferreira and J.A. Helayel-Neto, Phys.Rev. D \textbf{71}, 044018 (2005); J.L.
Boldo, J.A. Helayel-Neto, L.M. de Moraes, C.A.G. Sasaki and V.J. V. Otoya, Phys.
Lett. B \textbf{689}, 112 (2010).

\bibitem {Gravity2}V. A. Kostelecky and J. D. Tasson, Phys. Rev. D
\textbf{83}, 016013 (2011).

\bibitem {Jackiw}S.M. Carroll, G.B. Field, and R. Jackiw, Phys. Rev.\textit{\ }%
D \textbf{41}, 1231 (1990).

\bibitem {KM1}V. A. Kostelecky and M. Mewes, Phys. Rev. Lett. \textbf{87},
251304 (2001).

\bibitem {KM2}V. A. Kostelecky and M. Mewes, Phys. Rev. D\textbf{\ 66}, 056005
(2002); V. A. Kostelecky and M. Mewes, Phys. Rev. Lett. \textbf{97}, 140401 (2006).

\bibitem {Cherenkov2}B. Altschul, Nucl. Phys. B\textbf{796}, 262 (2008); B.
Altschul, Phys. Rev. Lett. \textbf{98}, 041603 (2007); C. Kaufhold and F.R.
Klinkhamer, Phys. Rev. D \textbf{76}, 025024 (2007).

\bibitem {Klink2}F.R. Klinkhamer and M. Risse, Phys. Rev. D \textbf{77},
016002 (2008); F.R. Klinkhamer and M. Risse, Phys. Rev. D \textbf{77}, 117901 (2008).

\bibitem {Klink3}F. R. Klinkhamer and M. Schreck, Phys. Rev. D \textbf{78},
085026 (2008).

\bibitem {Qasem}Q. Exirifard, Phys. Lett. B \textbf{699}, 1 (2011).

\bibitem {Kostelec}V. A. Kostelecky and M. Mewes, Phys. Rev.\textit{\ }D
\textbf{80}, 015020\ (2009); M. Cambiaso, R. Lehnert, and R. Potting, Phys.Rev. D
\textbf{85 }085023 (2012); M. Mewes, Phys. Rev. D\textbf{ 85}, 116012 (2012).

\bibitem {Defects}M.N. Barreto, D. Bazeia, and R. Menezes, Phys.Rev. D\textbf{73}, 065015 (2006); A. de Souza Dutra, M. Hott, and F. A. Barone, Phys. Rev. D
\textbf{74,} 085030 (2006); D. Bazeia, M. M. Ferreira Jr., A. R. Gomes, and R.
Menezes, Physica D (Amsterdam) \textbf{239}, 942 (2010); A. de Souza Dutra, and R. A. C.
Correa, Phys. Rev. D\textbf{\ 83}, 105007 (2011).

\bibitem {Monopole1}N.M. Barraz Jr., J.M. Fonseca, W.A. Moura-Melo, and J.A. Helay\"{e}l-Neto, Phys.Rev. D \textbf{76}, 027701 (2007); A. P. Baeta
Scarpelli and J. A. Helayel-Neto, Phys.Rev. D \textbf{73}, 105020 (2006).

\bibitem {Monopole2}M.D. Seifert, Phys. Rev. Lett. \textbf{105}, 201601 (2010).

\bibitem {Seifert}M.D. Seifert, Phys.Rev. D \textbf{82}, 125015 (2010).

\bibitem {Baeta2}A. P. Baeta Scarpelli, H. Belich, J. L. Boldo, and J. A.
Helayel-Neto, Phys.Rev. D\textbf{\ 67}, 085021 (2003).

\bibitem {Gamboa}H. Falomir, J. Gamboa, J. Lopez-Sarrion, F. Mendez, and A. J. da
Silva, Phys.Lett. B \textbf{632}, 740 (2006); Phys. Rev. D \textbf{74,} 047701 (2006).

\bibitem {Altschul}B. Altschul, Phys. Rev. Lett. \textbf{98}, 041603 (2007).

\bibitem {Paulo1}R. Casana, M. M. Ferreira Jr., A. R. Gomes, and P.R. D. Pinheiro,
Eur. Phys. J. C \textbf{62}, 573 (2009); Q. G. Bailey and V. A. Kostelecky,
Phys. Rev. D \textbf{70}, 076006 (2004).

\bibitem {Dielectric}J. Lee and S. Nam, Phys. Lett. \textbf{B 261}, 437
(1991); D. Bazeia, Phys. Rev. \textbf{D 46}, 1879 (1992).

\bibitem {PRL}E. Babaev, J. J\"{a}ykk\"{a}, M. Speight, Egor Babaev, Phys.
Rev. Lett. \textbf{103}, 237002 (2009); M. A. Silaev, Phys. Rev. \textbf{B}
\textbf{83}, 144519 (2011).

\bibitem {Carlisson2}R. Casana, M. M. Ferreira Jr., E. da
Hora, and C. Miller work under development.
\end{thebibliography}
\end{document}